\documentclass[structabstract]{aa}  

\usepackage[tbtags,fleqn]{amsmath}
\usepackage{amsfonts}
\usepackage{graphicx}
\usepackage{natbib}
\usepackage{url}
\usepackage{amssymb}
\usepackage{deluxetable}

\begin{document}

   \title{Orion revisited}

   \subtitle{II. The foreground population to Orion A}
   \titlerunning{The foreground population to Orion A}
   \author{H. Bouy\inst{1}
          \and J. Alves\inst{2}
          \and E. Bertin\inst{3}
          \and L.~M. Sarro\inst{4}
          \and D. Barrado\inst{1}
          }

   \institute{Centro de Astrobiolog\'{i}a, INTA-CSIC, Depto Astrof\'\i sica, ESAC Campus, PO Box 78, 28691 Villanueva de la Ca\~nada, Madrid, Spain \\
             \email{hbouy@cab.inta-csic.es}
             \and
             Department of Astrophysics, University of Vienna, T\"urkenschanzstrasse 17, 1180 Vienna, Austria\\
             \and
             Institut d'Astrophysique de Paris, CNRS UMR 7095 UPMC, 98bis bd Arago, F-75014 Paris, France \\
             \and 
             Dpto. de Inteligencia Artificial, ETSI Inform\'atica, UNED, Juan del Rosal, 16, E-28040, Madrid, Spain\\
             }

   \date{Received ....; Accepted 03/02/2014} 


  \abstract{}
  {Following the recent discovery of a large population of young stars in front of the Orion Nebula, we carried out an observational campaign with the \emph{DECam} wide-field camera covering $\approx$10~deg$^{2}$ centered on NGC 1980 to confirm, probe the extent of, and characterize this foreground population of pre-main-sequence stars.}
  {We used  multiwavelength wide-field images and catalogs to identify potential foreground pre-main-sequence stars using a novel probabilistic technique based on a careful selection of colors and luminosities.}
  {We confirm the presence of a large foreground population towards the Orion A cloud. This population contains several distinct subgroups, including NGC~1980 and NGC~1981, and stretches across several degrees in front of the Orion~A cloud. By comparing the location of their sequence in various color-magnitude diagrams with other clusters, we found a distance and an age of 380~pc and 5$\sim$10~Myr, in good agreement with previous estimates. Our final sample includes 2123 candidate members and is complete from below the hydrogen-burning limit to about 0.3~M$_{\sun}$ , where the data start to be limited by saturation. Extrapolating the mass function to the high masses, we estimate a total number of $\approx$2\,600 members in the surveyed region.  }
 {We confirm the presence of a rich, contiguous, and essentially coeval population of about 2600 foreground stars in front of the Orion~A cloud, loosely clustered around NGC~1980,  NGC~1981, and a new group in the foreground of the OMC-2/3. For the area of the cloud surveyed, this result implies that there are more young stars in the foreground population than young stars inside the cloud. Assuming a normal initial mass function,  we estimate that between one to a few supernovae must have exploded in the foreground population in the past few million years, close to the surface of Orion~A, which might be responsible, together with stellar winds, for the structure and star formation activity in these clouds. This long-overlooked foreground stellar population is of great significance, calling for a revision of the star formation history in this region of the Galaxy.}


\maketitle

%

\section{Introduction}
\label{sec:introduction}
The Orion star-forming complex is the nearest active star-forming region to Earth that produces massive stars, and is one of the richest in the nearest 1-kpc. It has long been recognized as a benchmark laboratory for star and planet formation studies as well as for the formation and dispersal of OB associations. The entire Orion star formation complex spreads across 200 pc and has spawned about 10$^4$ stars in the past 12 Myr \citep[e.g.][]{1994A&A...289..101B,2008hsf1.book..459B,2008hsf1.book..483M,2008hsf1.book..838B}. There are at least four subpopulations of young stars within the complex, first identified by \citet{1964ARA&A...2..213B}, from the older and dust-free Orion OB 1a group in the North (age $\sim$10 Myr) to the still-embedded Orion OB 1d (age $\sim$1 Myr), or the Orion Nebula Cluster, to the South. It was realized
early that the groups Orion OB 1c and 1d overlapped, at least in part, along the line of sight \citep[e.g.][]{1978ApJS...36..497W,1998AJ....115.1524G}.

Recently, \citet{2012A&A...547A..97A} used the denser regions of the Orion A cloud to block the background light in the optical, effectively isolating the stellar population in front of it, and found a rich stellar population in front of this cloud. The surprising result in this paper was not the confirmation of the existence of a foreground population, but how rich this population is, in particular towards the previously unstudied NGC 1980 cluster, which was shown to be slightly older ($\sim$ 5 Myr) than the embedded population in the cloud and contains a full stellar spectrum from OB stars to substellar objects. Using X-ray observations, \citet{2013ApJ...768...99P} recently found a population of sources towards the Orion A cloud that shows a very low amount of extinction (on the order of N$_{H} = 3\times10^{20}$ cm$^{-2}$, or about 0.1~mag of visual extinction) towards $\iota-$Ori and NGC~1980. This very low extinction population coincides with the foreground population isolated in \citet{2012A&A...547A..97A}, further confirming its presence.  The existence of a rich foreground population implies that the closest massive star formation and cluster formation benchmark, namely the Orion Nebula Cluster \citep[e.g.][]{2012ApJ...748...14D}, suffers from significant foreground contamination (on the order of 10\% for the Orion Nebula Cluster, but higher for the less rich L1641 young population). This result also implies that massive star-forming regions  can have complex star formation histories, and superposition of distinct young populations along a given line of sight probably is the norm, which complicates the determinations of, for example, the stellar yield of a molecular cloud, the duration of the star formation process (age spreads), and mass functions.   
 
\citet{2012A&A...547A..97A} made a first estimation of the size of the population (about 2000 objects) by extrapolating from the relatively high extinction regions of Orion A to a larger field.  This estimate was necessarily approximate, and called for  a better characterization of the foreground population. In the current paper we make use of data from a dedicated observational campaign with the DECam wide-field camera, covering about 10 square degrees of sky centered on NGC~1980, to confirm, probe the extent of, and characterize the foreground population. 

This paper is structured as follows: in Sect. 2 we describe the observational data acquired for this project as well as the archival data used. In Sect. 3 we present the results of our approach, namely the identification of the two foreground populations and their characterization. We present a general discussion on the importance of the result found in Sect. 4 and summarize our main results in Sect. 5.

\section{Data}
\label{sec:data}

To characterize the foreground population to the Orion A molecular cloud we made use of existing surveys together with raw data from Cerro Tololo Inter-American Observatory (CTIO) and its new wide-field Dark Energy Camera ({\it DECam}).

\subsection{CTIO/DECam}
\label{sec:ctiodecam}
\begin{figure}
   \centering
   \includegraphics[width=0.45\textwidth]{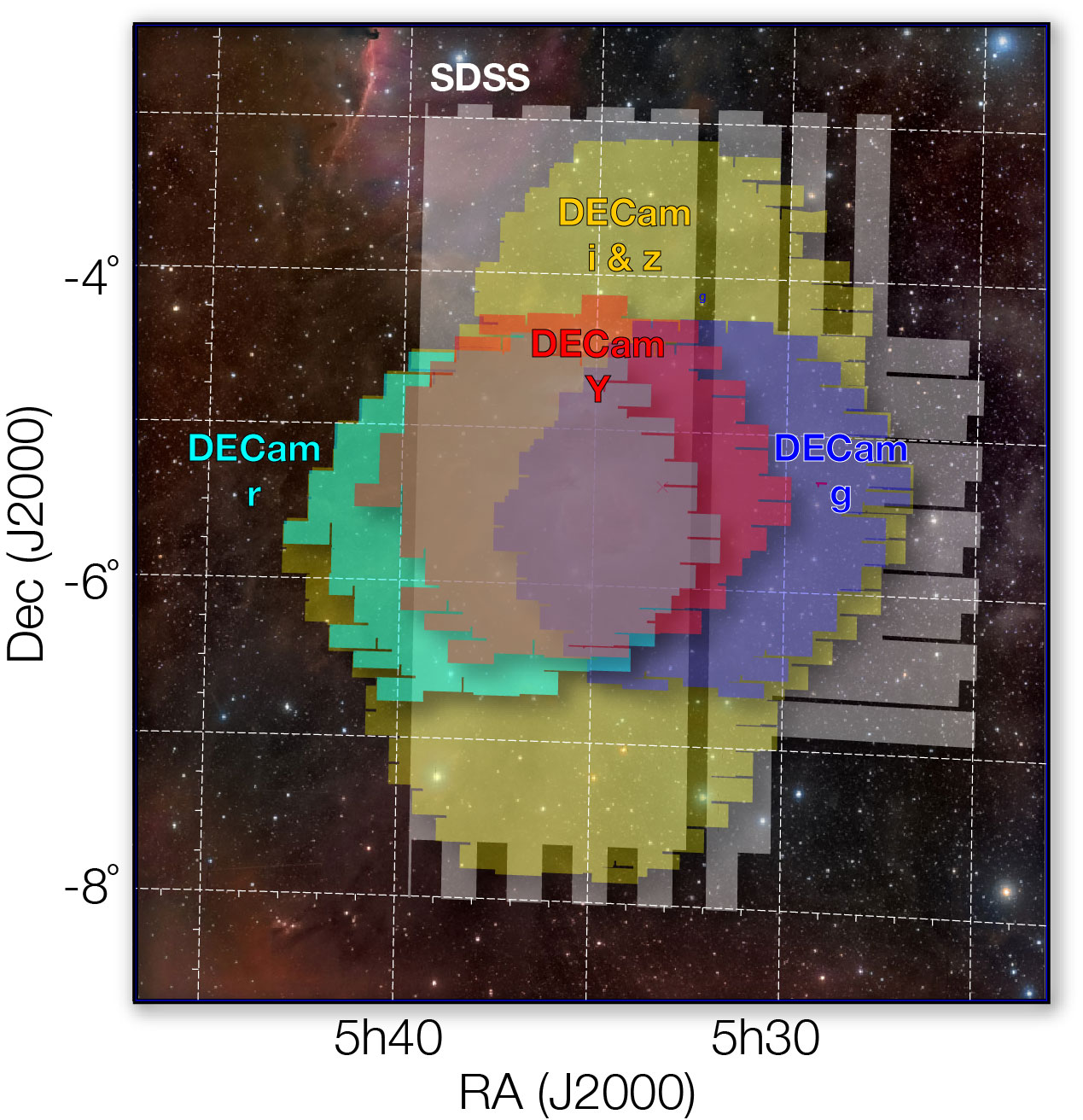}
      \caption{Coverage of the SDSS and {\it DECam} images used in this study. The white area corresponds to the SDSS survey. The blue, cyan and red areas correspond to the {\it DECam} $g$, $r$ and $Y$-band observations, respectively. The yellow area was observed with {\it DECam} in the $i$ and $z$ bands. Background photograph courtesy of Rogelio Bernal Andreo (DeepSkyColors.com)}
         \label{fig:coverage}
   \end{figure}

The Orion A cloud was observed at the Cerro Tololo Inter-American Observatory (CTIO) with the {\it DECam} wide-field camera on the Blanco telescope in the Sloan $griz$ and $Y$ filters on 2012 October 30 and November 19, 20, and 22  (P.I. Bertin \& Bouy) during the science verification of the instrument. Figure~\ref{fig:coverage} gives an overview of the area covered by these observations. The conditions were clear. The full width at half maximum (FWHM) measured in the images oscillated between 0.9--2\farcs3. These numbers are indicative only because the point spread function (PSF) was sometimes very elongated and highly asymmetric. DECam images where indeed obtained while the instrument was still being tested, and suffered from tracking problems in some parts of the sky, leading to variable amounts of PSF elongation and distortions. This problem was eventually fixed soon after our observations were completed. Sets of short, intermediate, and long exposures were obtained in each filter, as described in Table~\ref{tab:obs_decam}. The individual raw images were processed using an updated version of \emph{Alambic} \citep{2002SPIE.4847..123V}, a software suite developed and optimized for the processing of large multi-CCD imagers, and adapted for {\it DECam}. \emph{Alambic} includes standard processing procedures such as overscan and bias subtraction for each individual readout port of each CCD, flat-field correction, bad-pixel masking, and CCD-to-CCD gain harmonization. {\it DECam} is only moderately affected by fringes even in the reddest filters, with fringe amplitudes $<$1\% (DES consortium, priv. communication), and no fringe correction was attempted here. Aperture and PSF photometry were extracted from the individual images using {\sc SExtractor} \citep{1996A&AS..117..393B} and {\sc PSFEx} \citep{PSFEx}. The individual catalogs were then registered and aligned on the same photometric scale using {\sc Scamp} \citep{2006ASPC..351..112B}. The photometric zero-points of the merged catalog were derived by cross-matching with the  {\it Sloan Digital Sky Survey III} DR9 catalog \citep[SDSS DR9][]{SDSSDR9} for the $griz$ filters, and with the UKIDSS catalog \citep{UKIDSS} for the $Y$ filter. The {\it DECam} observations typically reach a 3-$\sigma$ completeness limit of 23$\sim$24~mag and complement the SDSS and UKIDSS catalogs in their faint end and around the bright Orion Nebula region that is lacking in both the SDSS and UKIDSS data. 

\begin{table}
\caption{{\it CTIO/DECam} observations\label{tab:obs_decam}}
\begin{tabular}{lcc}\hline\hline
Filter          &  Exposure time (s)            \\
\hline
$g$   & 3,90 \\
$r$    & 20, 90 \\ 
$i$    & 3, 30, 90 \\
$z$   & 3, 30, 90 \\
$Y$   & 30 \\
\hline
\end{tabular}
\end{table}

\subsection{Public catalogs}
\label{sec:catalogues}
We retrieved the astrometry and photometry for all sources within a box slightly larger than the {\it DECam} survey and encompassing the region between  82\fdg6$<$RA$<$85\fdg0 and -7\fdg6$<$Dec$<$-3\fdg8 in the SDSS DR9 catalog, 2MASS catalog \citep{2MASS}, UKIDSS catalog, and APASS catalog \citep{2009AAS...21440702H}. Table~\ref{tab:catalogues} gives a summary of the filters used for these catalogs.

\begin{table}
\caption{Catalogs and observations used in this study}
 \label{tab:catalogues}
\begin{tabular}{lc}\hline\hline
 Instrument or survey    &  Band    \\
         &                       \\
\hline
SDSS (DR9)           & $u$,$g$,$r$,$i$,$z$   \\
CTIO/DECam  & $g$,$r$,$i$,$z$,$Y$     \\
2MASS           & $J$,$H$,$Ks$ \\
UKIDSS (DR10) & $Z$,$Y$,$J$,$H$,$Ks$ \\
APASS  (DR1)  & $B$,$V$,$r$,$i$ \\
\hline
\end{tabular}
\end{table}

\section{Results: a pre-main-sequence population in front of the Orion A association\label{selection}}
\label{sec:results}

\subsection{General considerations}
\begin{figure}
   \centering
   \includegraphics[width=0.49\textwidth]{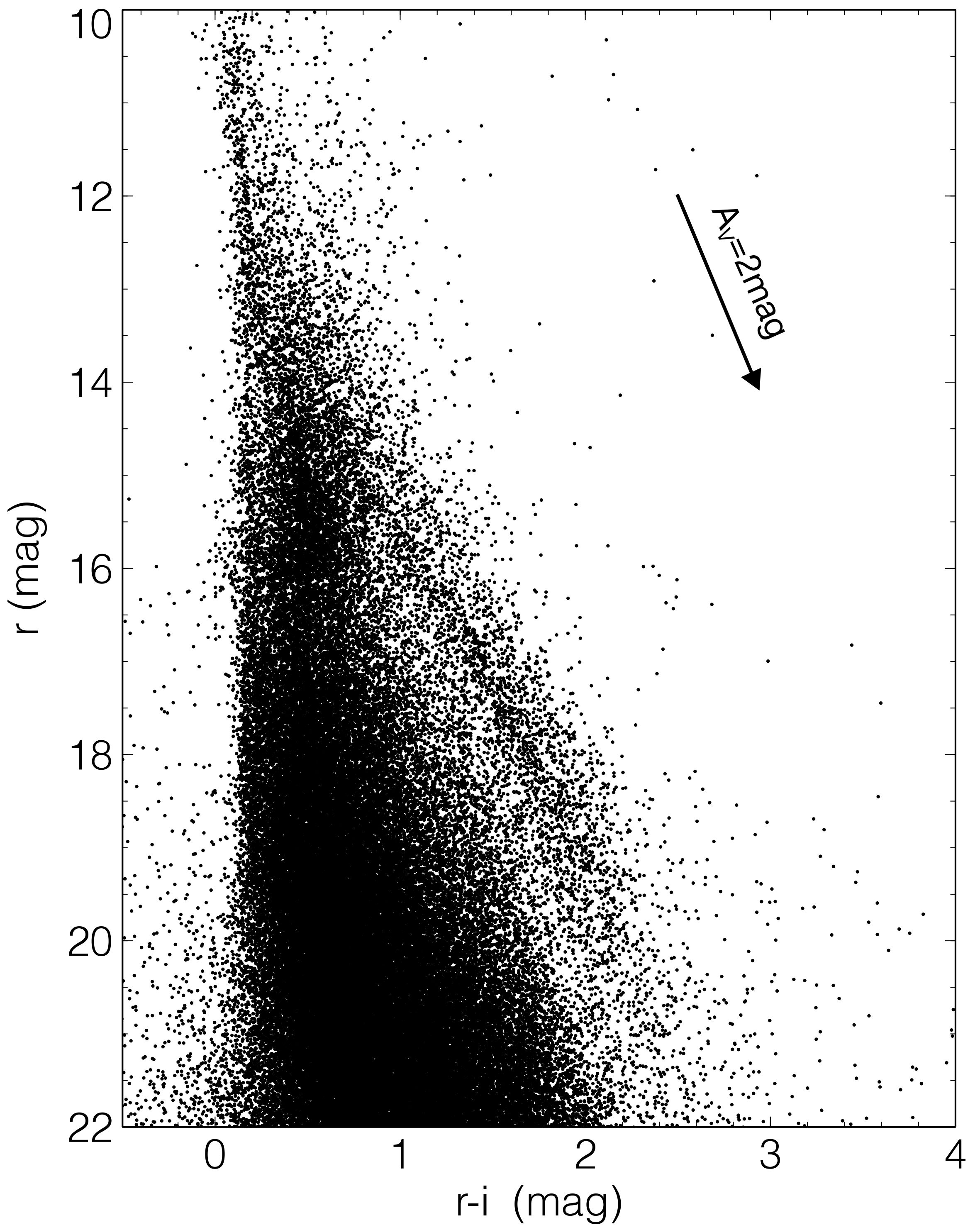}
      \caption{$r$ vs $r-i$  color-magnitude diagram including the {\it DECam} measurements complemented by SDSS and APASS. An arrow represents an extinction vector of 2~mag.}
         \label{fig:r_rmi}
   \end{figure}

Figure~\ref{fig:r_rmi} shows a $r$ vs $r-i$ color-magnitude diagram of the region. A dense redder sequence is immediately identifiable, indicating the presence of a rich co-eval population. The sequence is also identifiable in any other color-magnitude diagrams that we drew from our dataset. Several basic characteristics can be inferred from this diagram immediately:
\begin{itemize}
\item the density of the sequence suggests that the group is rich and massive,
\item the sequence appears to be mostly unaffected by reddening, indicating that the group is in front of the Orion A cloud and associations,
\item the dispersion of the sequence (less than $\approx$1~mag) is lower than typically seen for very young ($<$5~Myr) cluster \citep[][]{2007MNRAS.375.1220M}.
\end{itemize}

\subsection{Member selection}
\begin{figure}
   \centering
   \includegraphics[width=0.40\textwidth]{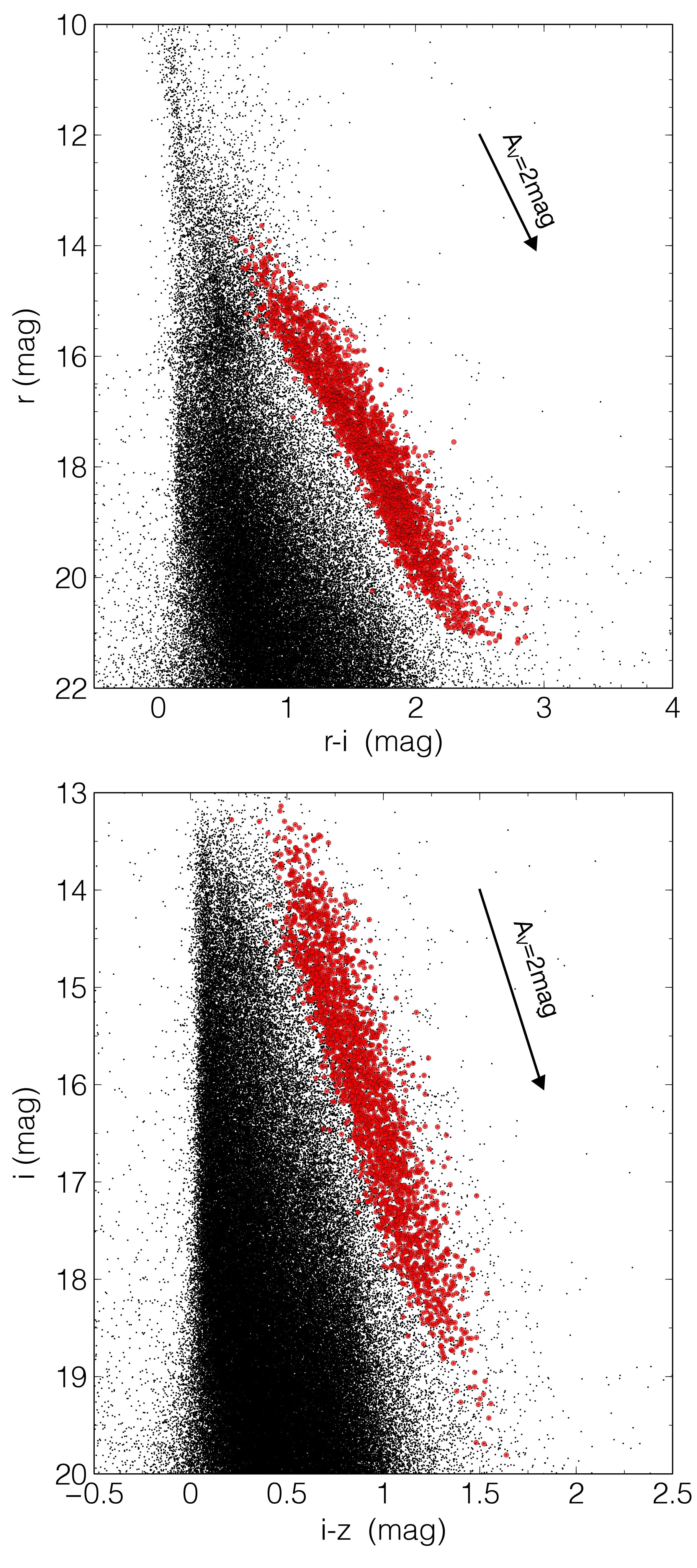}
      \caption{$r$ vs $r-i$ (upper panel) and $i$ vs $i-z$ (lower panel) color-magnitude diagrams of the survey. Objects with a membership probability $\ge$99.5\% are represented with red dots.}
         \label{fig:sequence}
   \end{figure}

Cluster members are traditionally identified based on their location in various color-magnitude diagrams with respect to theoretical or empirical isochrones. This method suffers from several limitations. Because the color-magnitude diagrams are analyzed sequentially, any member with partial photometric measurements will be missed. Additionally, this method does not quantify the membership probability of a given source. In this study, we selected members using a novel multidimensional probabilistic analysis using carefully selected color-magnitude diagrams and luminosities. The method is described in detail in \citet{luisma}. Briefly, an initial training set is obtained through the following steps. First a simple {\it ad hoc} cut in the $r$ versus $r-i$ diagram of sources (with measurements available in all the $g,r,i,z,Y,H$,and $K$ bands) yields a subset of sources with a probability density dominated by cluster members. Then, a multivariate Gaussian distribution is fit to this subset and sources at Mahalanobis\footnote{The Mahalanobis distance provides a relative measure of a data point's distance from a reference in a multidimensional space.} distances (from its mean) greater than three are removed from the set. Finally, a principal curve \citep{Principal_Curves} is fit to the resulting dataset in the multidimensional color-magnitude space spanned by $r,z,H,(r-i),(i-K)$, and $(g-i)$.  At the same time, a mixture-of-Gaussians model is fit to the remaining set of field/background sources, comprising those below the linear cut in the $r$ vs. $(r-i)$ diagram and those at Mahalanobis distances above three (in the same multidimensional color-magnitude space). The complexity of the mixture-of-Gaussians model is determined by the optimal Bayesian information criterion (BIC) value. From that point onward, an expectation-maximization iterative scheme is applied, whereby in each cycle we
\begin{enumerate}
\item[i)] first compute the expected membership probability of sources according to these two models (mixture-of-Gaussians for the field and a principal curve for the cluster members), and then
\item[ii)] we modify the membership list accordingly (removing low-probability sources and adding high-probability ones), and infer a new principal curve model that accounts for the changes. 
\end{enumerate}
New high-probability members (defined here conservatively as having a probability $\ge$99.75\%) are then selected as the new training set, while originally selected members with a lower probability are rejected from the training set. 
The calculation takes into account measurement uncertainties on the photometry, as well as data censoring. This is particularly important since the various datasets combined in our study have very different coverage (Fig.~\ref{fig:coverage}) and depths (Table~\ref{tab:obs_decam}). The method leads to 2\,123 high-confidence (probability$\ge$99.5\%) cluster members among the 605\,020 sources of our initial catalog. Figure~\ref{fig:sequence} illustrates the results and shows the sequence obtained in two color-magnitude diagrams. 

Members of coincident background young associations (such as those in the ONC) most likely contaminate the final sample. Assuming that the young population associated with the ONC and the Orion A cloud suffers from extinction higher than about A$_V\sim 1$~mag, a conservative assumption, a simple independent test on how severe this contamination might be is to search for reddened sources in a sensitive near-infrared (NIR) color-color diagram. We found that of the selected sample, only about 1--2\% of the sources showed NIR colors indicating A$_V> 1$~mag, which gives additional confidence to our selection procedure.  

Our selection process is, regardless, expected to give a lower contamination rate and a more complete list of members than the traditional method described above. As illustrated in Fig.~\ref{fig:sequence}, the selection method works best in the luminosity range where the cluster sequence is clearly separated from the background/foreground population. We expect the resulting sample to be mostly complete in the luminosity range between 15$\lesssim r\lesssim$19.5~mag and 14.5$\lesssim i\lesssim$17.5~mag.

\subsection{Spatial distribution and origin}
\begin{figure*}
   \centering
   \includegraphics[width=0.95\textwidth]{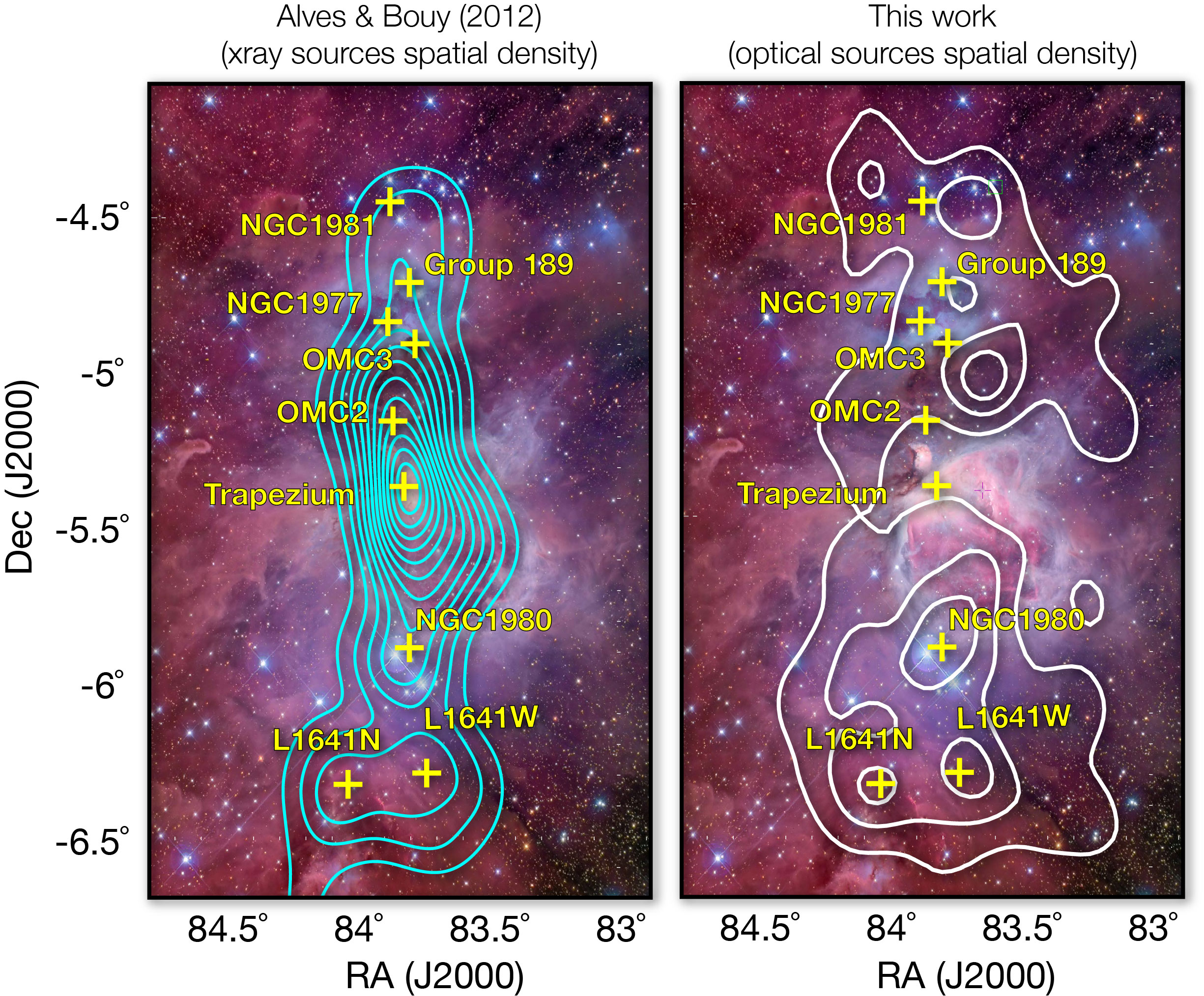}
      \caption{Left: Contours of the spatial density distribution of X-ray sources from \citet{2012A&A...547A..97A}. Right: Contours of spatial density distribution of optically selected candidate members of the foreground population. Background photograph courtesy of Jon Christensen (christensenastroimages.com)}
         \label{fig:spatial_distrib}
   \end{figure*}

Figure~\ref{fig:spatial_distrib} shows the spatial distribution of candidate members with a membership probability higher than 99.5\%  (hereafter OriA-foreground). The sources cluster around several known groups: NGC~1980, NGC~1981, L1641W, and L1641N. An overdensity is also visible southwest of OMC~3 on one hand, and at the position reported by \citet{2000ESASP.445..347C} for Group~189 in both X-ray and optical source density maps on the other hand.  

NGC~1980 and NGC~1981 belong to a foreground population, as demonstrated by \citet{2012A&A...547A..97A} and \citet{2010MNRAS.407.1875M}, respectively.  Although we cannot rule out that L1641W is associated to the Orion~A cloud, we note that both its projected distance away from the cloud and its age suggest that it belongs to the foreground population. The overdensity located on top of L1641N could be produced by genuine L1641N members. The low extinction towards these sources (as found in the NIR color-color diagram) and the small fraction of stars with mid-infrared excess nevertheless suggest that it could be a distinct population. It is not possible to draw firm conclusions and we consider the distance and origin of this group undetermined and to be confirmed.

\begin{figure}
   \centering
   \includegraphics[width=0.45\textwidth]{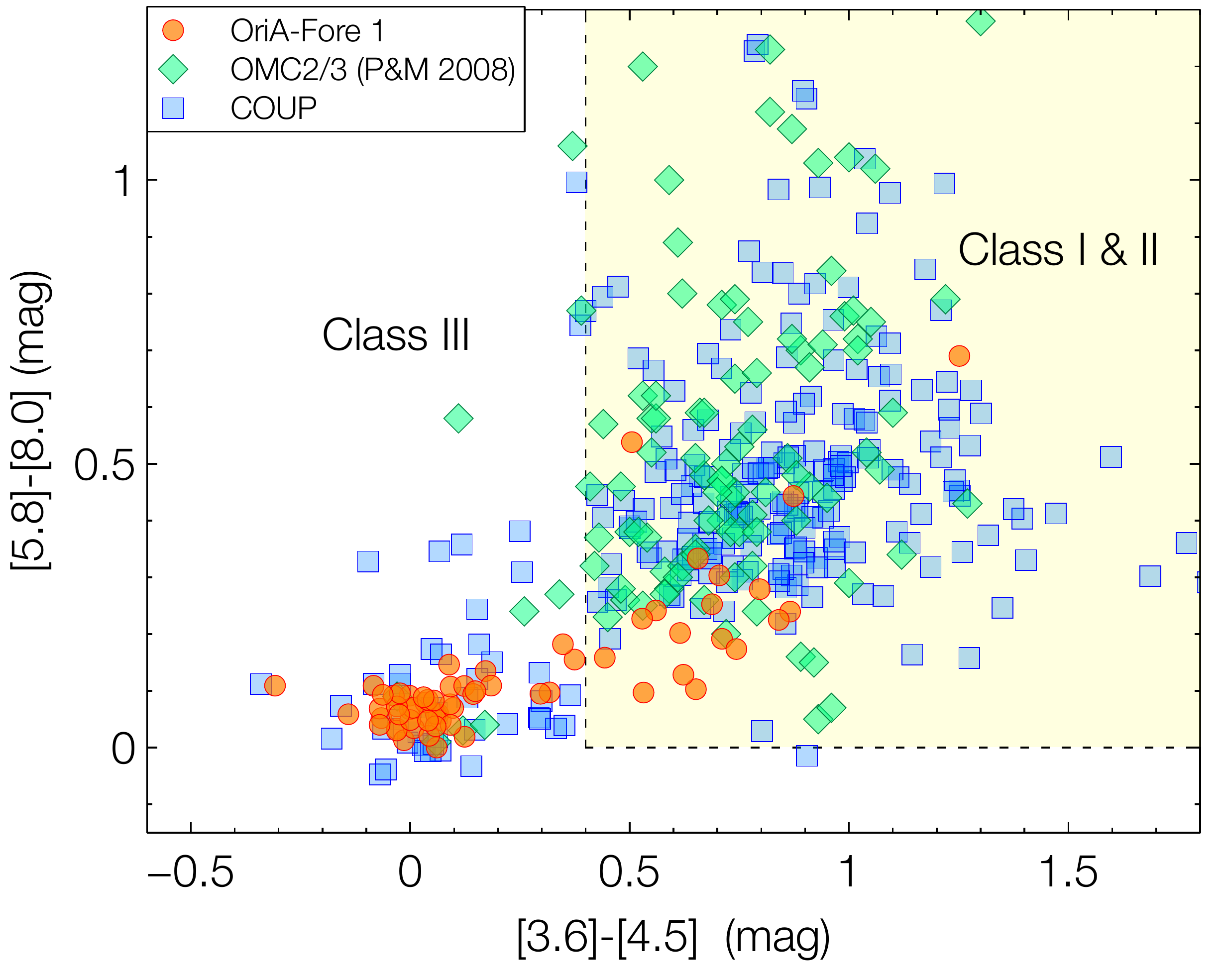}
      \caption{[3.6]-[4.5] vs [5.8]-[8.0] color-color diagram for OriA-Fore~1 (orange dots, sources located within 20\arcmin\, of the centroid of the overdensity, see Fig.~\ref{fig:spatial_distrib}), for known OMC2/3 members from \citet{2008hsf1.book..590P} (green diamonds) and for the sample of ONC members from the COUP survey (blue squares). Dotted lines represent the boundary between class~III and class I \& II sources (shaded area) as defined in \citet{2004ApJS..154..363A}. }
         \label{fig:omc3}
   \end{figure}

The overdensity of sources southwest of OMC3 does not correspond to any group previously identified in the literature, and we hereafter refer to it as OriA-Fore 1. It does not appear in the X-ray source density map of \citet{2012A&A...547A..97A} because it falls near the edge of the field covered by the {\it XMM-Newton} observations. Figure~\ref{fig:omc3} shows a {\it Spitzer} [3.6]-[4.5] vs [5.8]-[8.0] color-color diagram for sources located within 20\arcmin\, from the centroid of the overdensity. ONC and OMC2/3 members from the literature are also represented for comparison. This diagram produces an excellent diagnostic to easily distinguish between objects with and without disks. The vast majority of the OriA-Fore~1 sources show no or little excess, indicating a more advanced evolutionary status than their ONC or OMC2/3 counterparts, and suggesting that they are a distinct population. 

Interestingly, \citet{2012A&A...547A..97A} found a velocity dispersion for sources located in the vicinity of OriA-Fore~1 (see their Fig.~8) significantly different from those of the ONC and NGC~1980 (23.3$\pm$3.0~km s$^{-1}$ for the group that falls on top of the dense gas in OMC2 and 27.5$\pm$2.3~km s$^{-1}$ for the new group, to the West of OMC2). The errors are too large and the samples relatively small so that we cannot draw a strong conclusion, but both the spatial distribution and velocity dispersion are suggestive of two overlapping populations.  Approximately 70 sources are located within 10\arcmin\, of the centroid, and 162 within 20\arcmin. The absence of bright massive OB stars in this area and the low luminosity of the sources (14$\leq i\leq$18~mag) suggest that the group forms a low-mass cluster.  The current datasets do not allow us to draw a firm conclusion on its distance. Its location away from the cloud, the low disk-frequency, and the distinct velocity dispersion suggest that the group is not directly associated to the Orion~A. 

\begin{table*}
\centering
\caption{Orion groups and clusters}
 \label{tab:groups}
\begin{tabular}{lccccl}\hline\hline
Name                  &  RA   & Dec &   Age    & Distance   &    Reference \\
                           &  (J2000) & (J2000) & (Myr)  & (pc)           &                     \\
\hline
\multicolumn{6}{c}{Orion~A} \\
\hline
NGC~1980         &  83\fdg80 & -5\fd95 & 5$\sim$10 & $\sim$380                & (1)  \\
NGC~1981         &  83\fdg78 & -4\fd34 & 5$\pm$1  & 380$\pm$17 & (2) \\
Group~189         &  83\fdg84   & -4\fd71 & \ldots & \ldots & (3) \\
L1641N   & 83\fdg94  & -6\fd28 & 2$\sim$3 & \ldots & (4)  \\
Trapezium & 83\fdg84 & -5\fd40 & 1$\sim$3 & 390$\sim$440 & (4), (5), (6) \\
OMC~2 & 83\fdg86 & -5\fdg17 & $\lesssim$1 & 420$\sim$450 & (4), (6), (7) \\
OMC~3 & 83\fdg79 & -4\fdg92 & $\lesssim$1 & 420$\sim$450 & (4), (6), (7) \\
L1641W & 83\fdg69  & -6\fd28 & 5$\sim$10 & $\sim$380 ? & (1) \\
OriA-Fore~1 &  83\fdg62 &  -4\fd97 & 5$\sim$10 & $\sim$380 ?    & (1) \\
\hline
\multicolumn{6}{c}{Ori OB1abc \& $\lambda-$Ori} \\
\hline
Collinder~80 & 83\fdg88 & -1\fd10 & 1$\sim$7 & 385 & (8) \\
$\sigma-$Ori & 84\fdg69 & -2\fd60 & 3$\sim$5 & 385 & (8) \\
Collinder~69 & 83\fdg77 & +9\fd93 & 5$\sim$10 & 400 & (9) \\
25 Ori & 81\fdg19 & +1\fd85 & 7$\sim$10 & 380 & (10) \\
\hline
\end{tabular}

References: (1) This work; (2) \citet{2010MNRAS.407.1875M}; (3)
\citet{2000ESASP.445..347C}; (4) \citet{1998AJ....115.1524G}; (5)
\citet{2007PASJ...59..897H}; (6) \citet{2007A&A...474..515M}; (7)
\citet{2008hsf1.book..590P}; (8) \citet{2008A&A...485..931C}; (9) \citet{2007ApJ...664..481B}; (10) \citet{2007ApJ...661.1119B}
\end{table*}

\subsection{Age and distance}

Pre-main-sequence (hereafter PMS) fitting uses the positions of stars in color-magnitude diagrams to derive ages and distances to young star clusters. Instead of fitting the cluster's PMS with theoretical isochrones that can be uncertain at young ages \citep{2009ApJ...702L..27B}, we derived relative ages and distances by comparing the observed PMS with the empirical PMS of well-known clusters.

\subsubsection{Comparison of NGC~1980 and NGC~1981}
\begin{figure}
   \centering
   \includegraphics[width=0.45\textwidth]{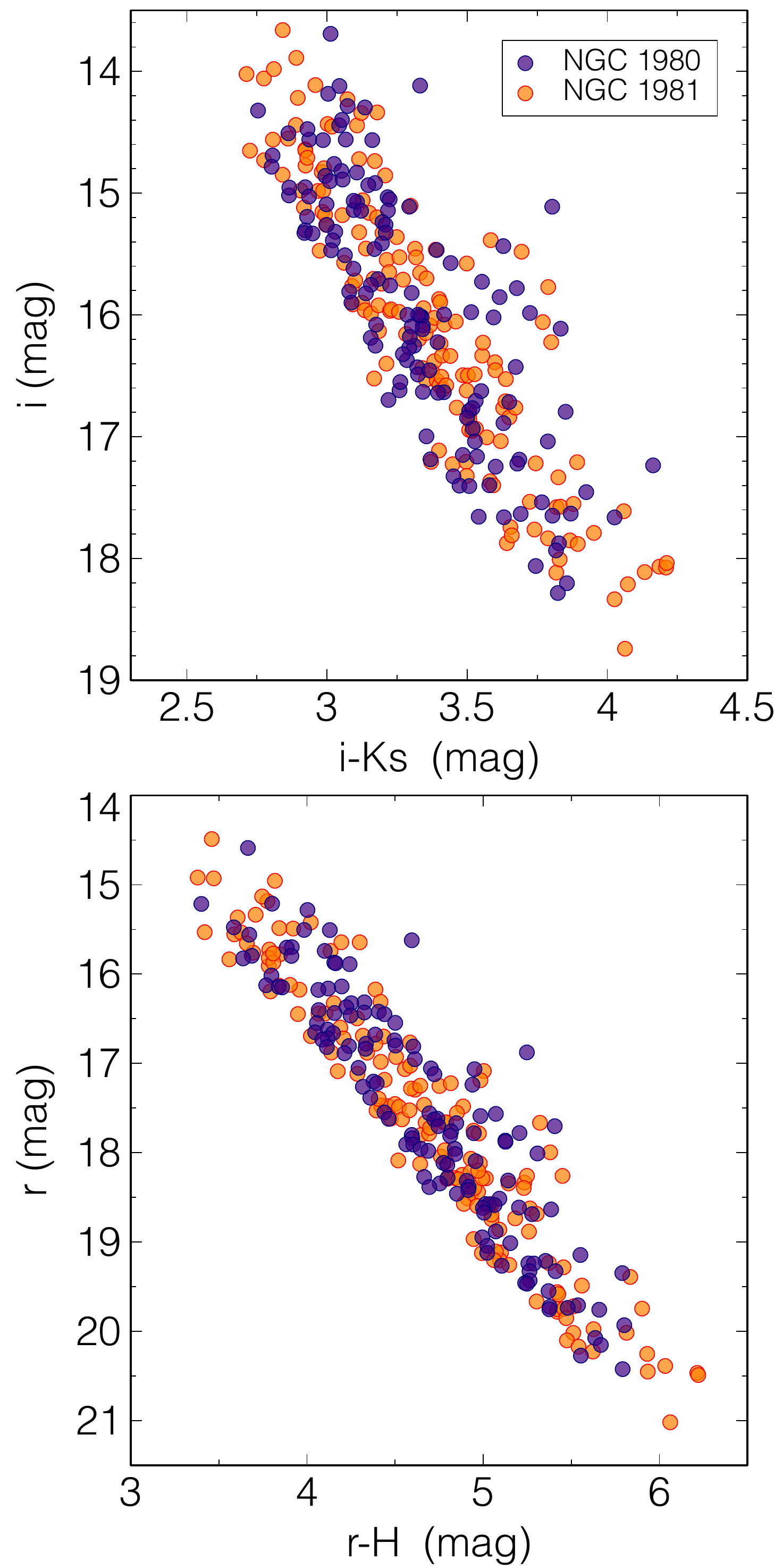}
      \caption{$r$ vs $(r-H)$ (lower panel) and a $i$ vs $(i-Ks)$ (upper panel) color-magnitude diagrams of sources selected within 20\arcmin\, of the NGC~1980 (violet) and NGC~1981 (orange) centers.}
         \label{fig:ngc1980_ngc1981}
   \end{figure}

To minimize the contamination by OMC~2/3, L1641N (and possibly L1641W and OriA-Fore~1, if they belong to Orion~A), we isolated high-probability members of NGC~1980 and NGC~1981 by selecting sources located within 20\arcmin\, of their respective centers (Table~\ref{tab:groups}). Figure~\ref{fig:ngc1980_ngc1981} shows that the two populations are indistinguishable in both a $r$ vs $(r-H)$ and a $i$ vs $(i-Ks)$ color-magnitude diagrams, suggesting that they have very similar ages and distances. \citet{2010MNRAS.407.1875M} reported a distance of 380$\pm$17~pc, a reddening of $E(B-V)$=0.07$\pm$0.03~mag, and an age of 5$\pm$1~Myr for NGC~1981, which places the foreground population 10$\sim$40~pc in front of the Orion~A cloud.

\subsubsection{Comparison with Collinder~69}
As an independent check, we compared the observed PMS with that of Collinder~69. Collinder~69 is a 5$\sim$10~Myr cluster located at $\approx$400~pc in the $\lambda-$Orionis star-forming region \citep{1999AJ....118.2409D, 2007ApJ...664..481B}. Collinder~69 presents several advantages for the purpose of our study. It is well studied and an extensive list of more than 200 spectroscopically confirmed members is available \citep{2012A&A...547A..80B}. Photometry in almost all the bands presented in the current study has been obtained for its members, making the comparison possible over a broad domain of colors and luminosities. Additionally, it is part of the $\lambda-$Orionis star-forming region, a subgroup of the Orion star formation complex (see Figure~\ref{fig:groups}).

\begin{figure}
   \centering
   \includegraphics[width=0.45\textwidth]{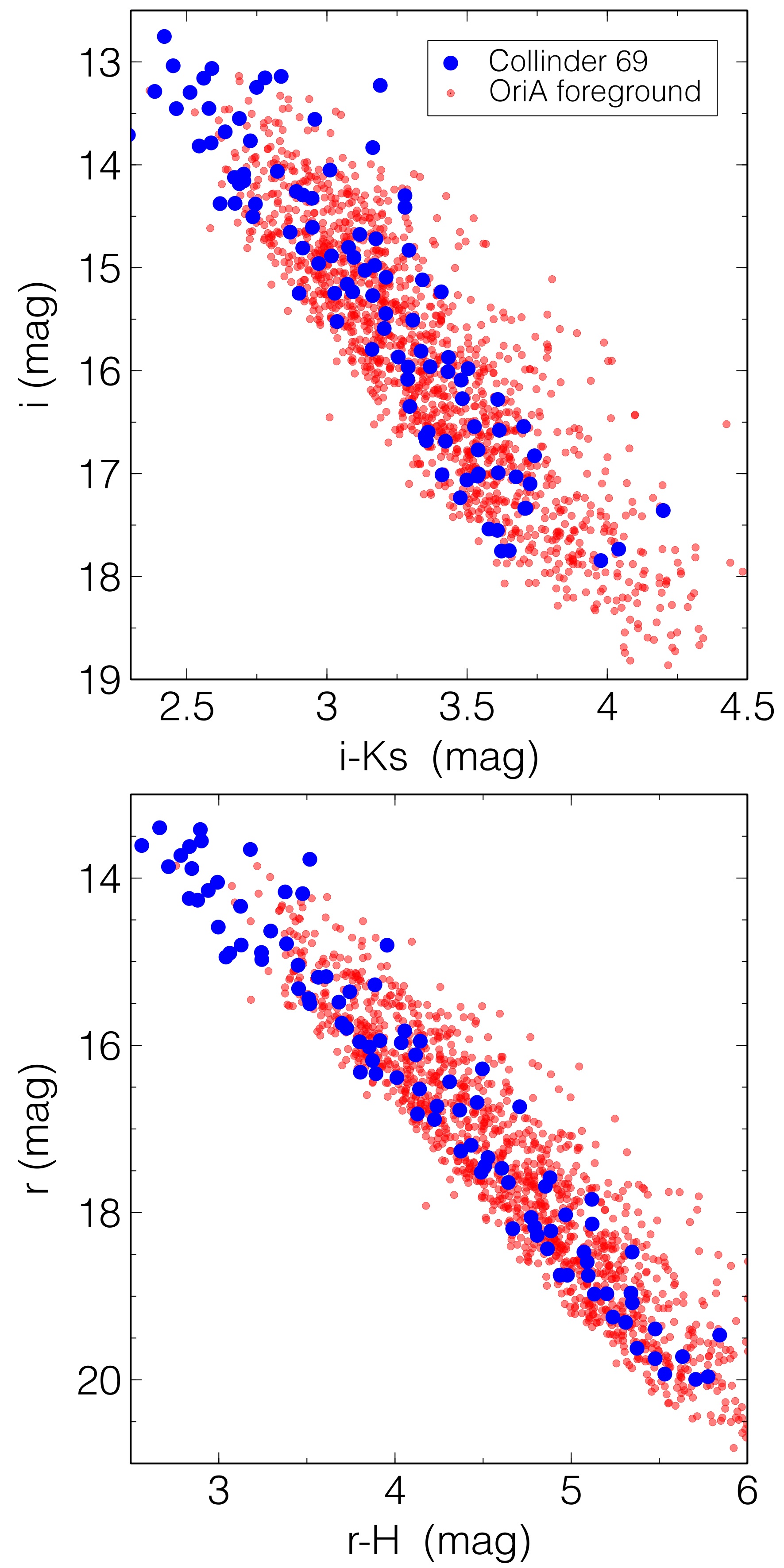}
      \caption{$r$ vs $(r-H)$ (lower panel) and a $i$ vs $(i-Ks)$ (upper panel) color-magnitude diagrams of OriA-foreground candidate members (red dots) with Collinder~69 confirmed members (blue dots).}
         \label{fig:cmd}
   \end{figure}

Figure~\ref{fig:cmd} shows a $r$ vs $(r-H)$ and a $i$ vs $(i-Ks)$ color-magnitude diagram of the candidate members selected previously with Collinder~69 confirmed members overplotted. Interstellar extinction towards Collinder~69 is relatively low \citep[A$_{\rm V}\approx$0.37~mag on average, ][]{1994ApJS...93..211D}, and we dereddened the photometry accordingly. Extinction towards OriA-foreground is not well established but is expected to be low since it lies in front of the Orion A molecular cloud \citep{2012A&A...547A..97A, 2013ApJ...768...99P,2010MNRAS.407.1875M}. The match is remarkably good in the two diagrams, especially considering that variability, excesses in the $riHKs$-bands (related to youth, accretion, rotation, and circumstellar disks), and extinction must affect the comparison. We conclude that NGC~1980 and NGC~1981 must have distances and ages similar to Collinder~69, which is consistent with the results of \citet{2010MNRAS.407.1875M}. The comparison of a cluster's PMS with an empirical PMS of differing age would indeed lead to a color-dependent distance modulus, which is not observed here. 

\subsubsection{Comparison with the ONC}
\begin{figure}
   \centering
   \includegraphics[width=0.45\textwidth]{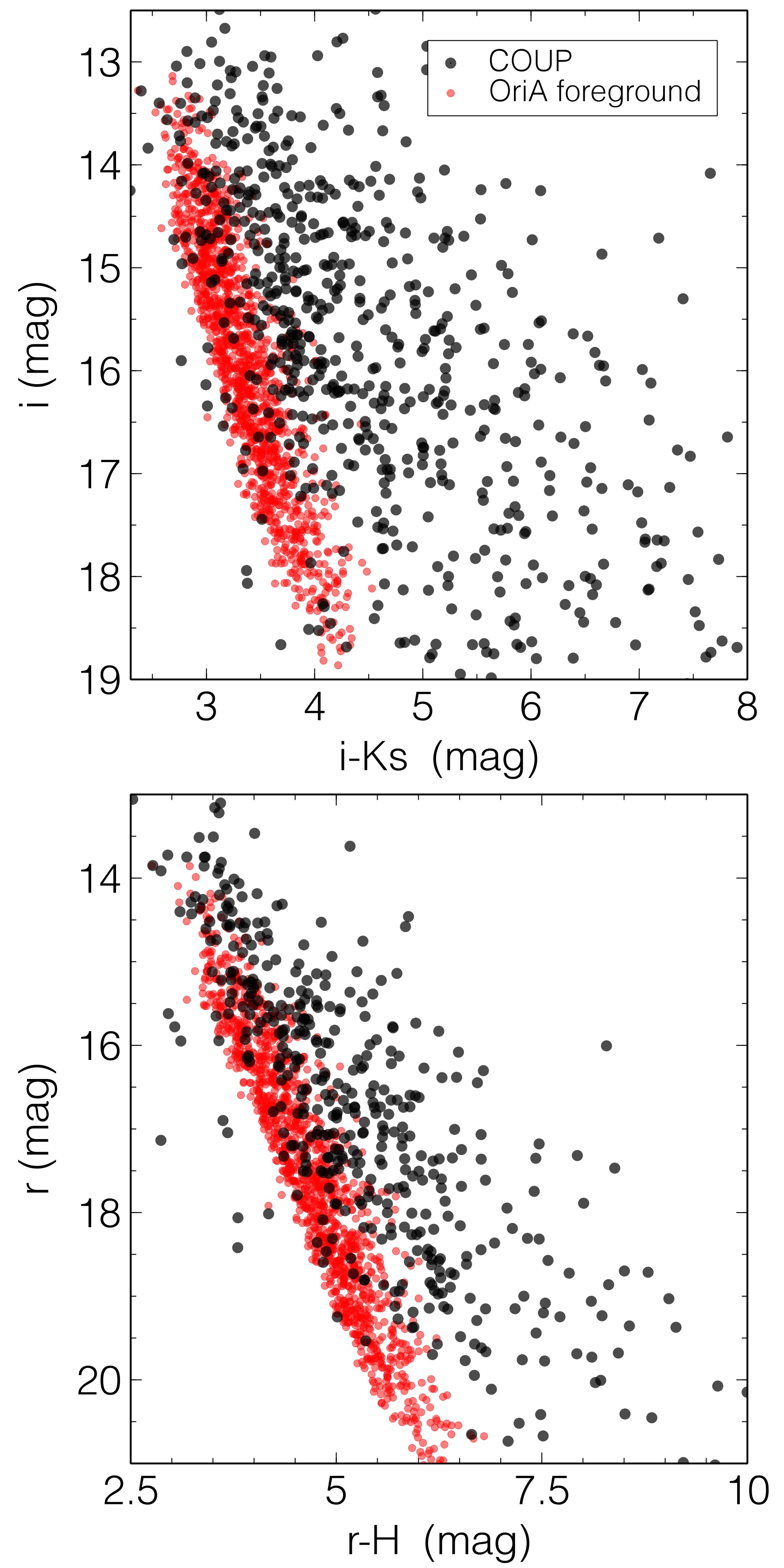}
      \caption{$r$ vs $(r-H)$ (lower panel) and a $i$ vs $(i-Ks)$ (upper panel) color-magnitude diagrams of OriA-foreground candidate members (red dots), and COUP members of the ONC (black dots).}
         \label{fig:cmd_coup}
   \end{figure}

Figure~\ref{fig:cmd_coup} compares the sequence formed by the OriA-foreground population in the same $r$ vs $(r-H)$ and a $i$ vs $(i-Ks)$ color-magnitude diagrams with the sequence formed by ONC members selected in the Chandra Orion Ultradeep Project \citep[COUP, ][]{2005ApJS..160..353G}. The two sequences are clearly different, the COUP sequence is on average brighter in spite of the higher and variable level of extinction that affects the members and the larger distance. The ONC sequence is also much more dispersed, a feature typical of very young ($<$5~Myr) clusters.  These differences can be interpreted as a difference in age and distance, the ONC members being significantly younger than the OriA-foreground population and embedded in the Orion~A cloud, in good agreement with the rest of our analysis.

For the rest of the study we consider a distance of 380~pc and an age of 5$\sim$10~Myr for OriA-foreground.

\subsection{Mass function}
\begin{figure}
   \centering
   \includegraphics[width=0.49\textwidth]{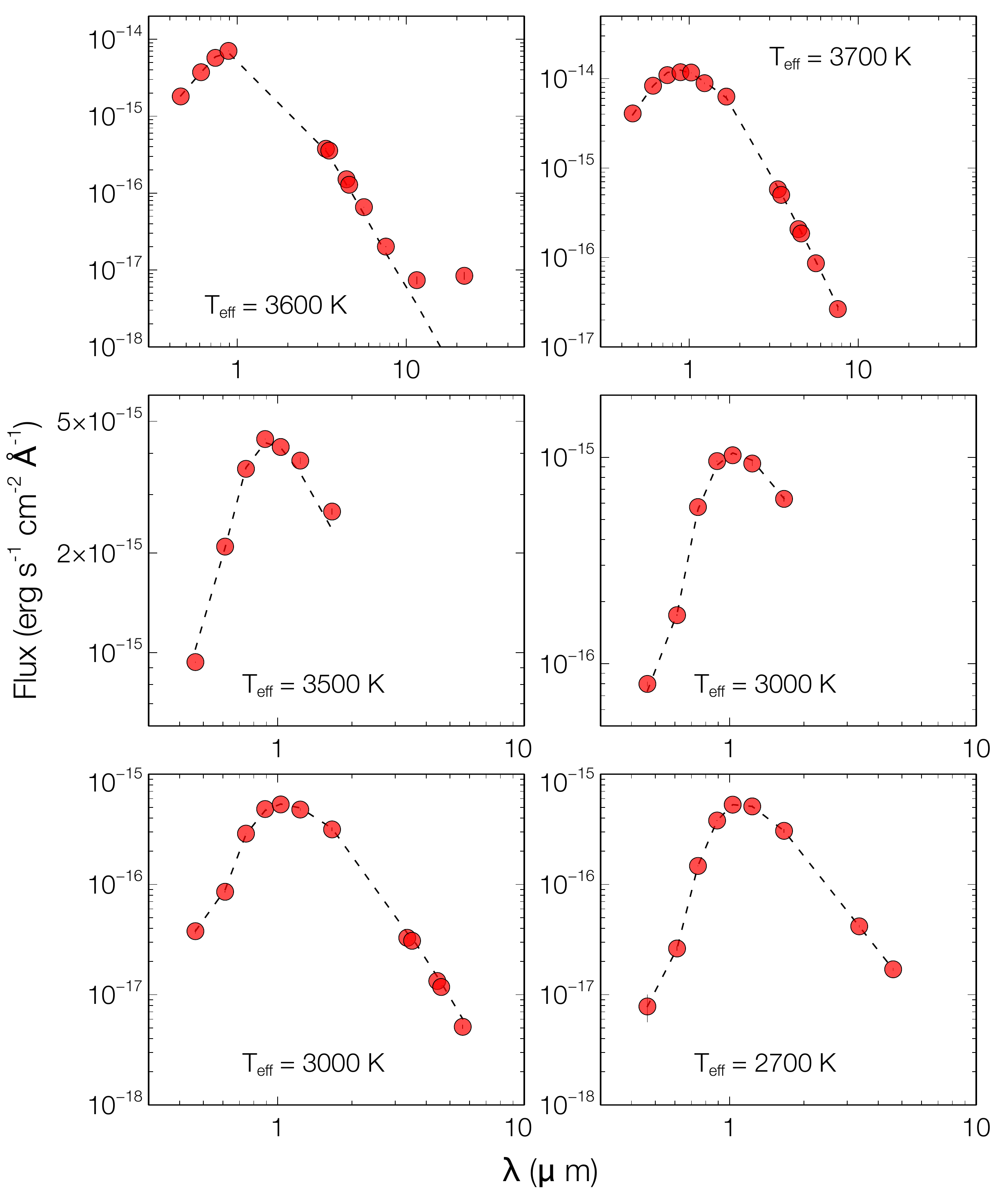}
      \caption{Examples of SED fitting obtained with VOSA. The red dots represent the observed photometry for a random sample of six sources selected as members. The dashed line represents the best-fit BT-Settl model. The corresponding T$_{\rm eff}$ is indicated. }
         \label{fig:seds}
   \end{figure}

We estimated the effective temperatures (hereafter T$_{\rm eff}$) of all OriA-foreground candidate members using the {\it virtual observatory SED analyzer} \citep[hereafter VOSA, ][]{2008A&A...492..277B}. VOSA offers the advantage of deriving robust T$_{\rm eff}$ independently of the distance and using all the available photometric information instead of a subset of colors and luminosities. The optical and near-infrared photometry of OriA-foreground candidate members described in Section~\ref{sec:data} was complemented with {\it Spitzer} \citep{2012A&A...547A..97A} and {\it WISE} \citep{2010AJ....140.1868W} mid-infrared photometry. Briefly, VOSA compares the observed spectral energy distributions (SED)  with a grid of  theoretical SEDs based on the BT-Settl models of \citet{2012EAS....57....3A}  and covering the range 2\,000$<$T$_{\rm eff}<$5\,000~K (by steps of 100~K) and 3.0$<log~g<$4.5 (by steps of 0.5~dex). As mentioned previously, extinction towards OriA-foreground is expected to be low and was neglected. VOSA automatically detects possible excesses from circumstellar disks (in the mid-infrared) or accretion (in the visible and UV), and rejects the corresponding measurements for the fit. The quality and quantity of the photometric measurements (between 4 and 14 measurements per source, with an average of 7 measurements per source) resulted in a robust fit for most sources, as illustrated in Fig.~\ref{fig:seds}. The T$_{\rm eff}$ were then transformed into masses using the (T$_{\rm eff}$ vs mass) relation given by the BT-Settl models assuming an age of 5~Myr. According to these calculations, the least-massive candidate members have masses of only $\approx$30~M$_{\rm Jup}$, independently of the distance. 

\begin{figure}
   \centering
   \includegraphics[width=0.49\textwidth]{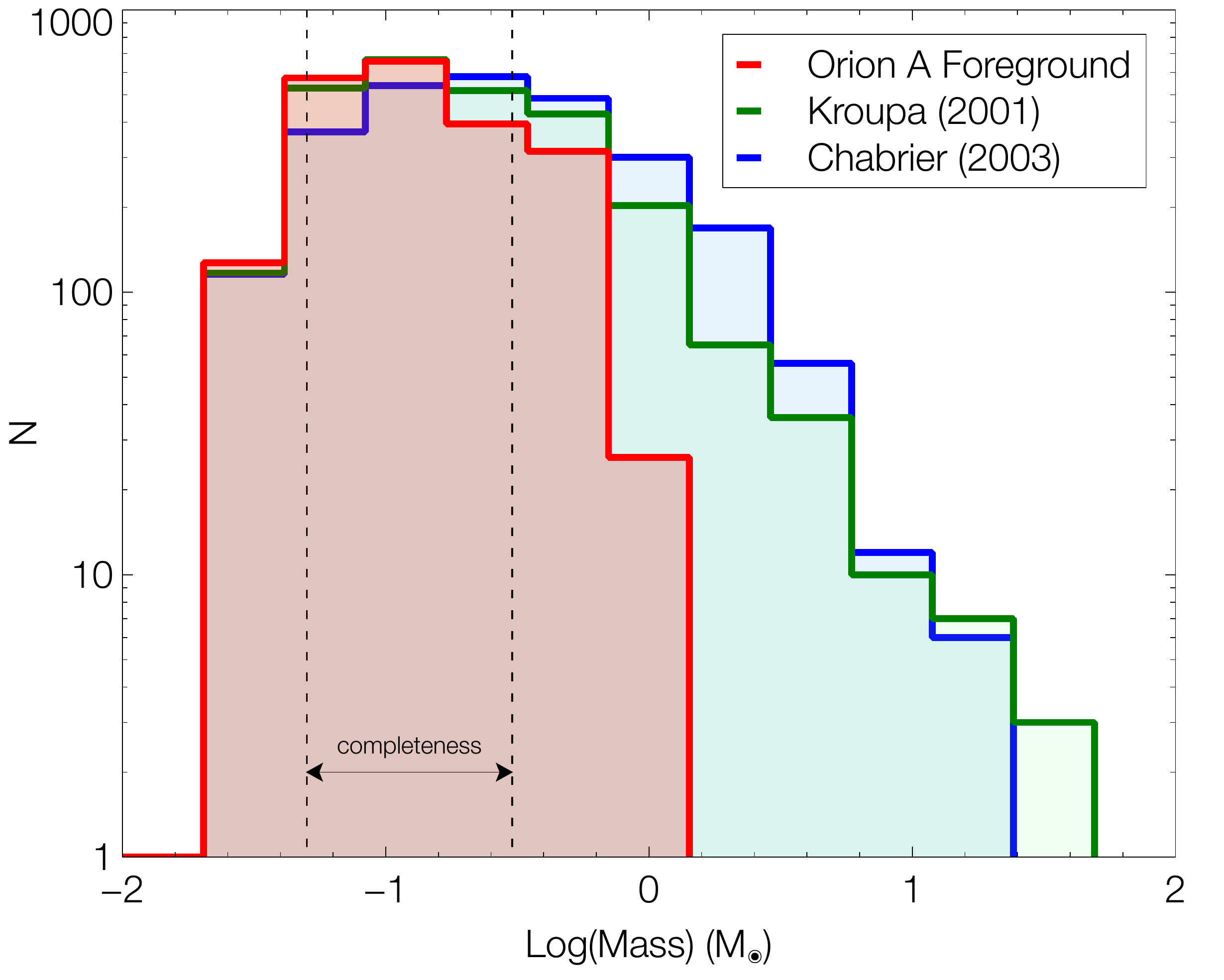}
      \caption{Mass function of OriA-foreground over the completeness domain computed for 5~Myr and 10~Myr. The predictions of \citet{2003PASP..115..763C} and \citet{2001MNRAS.322..231K} mass functions are overplotted.}
         \label{fig:imf}
   \end{figure}

The cluster's mass function was then computed over the luminosity range corresponding to the estimated completeness domain of the survey (described in Section~\ref{selection}). Figure~\ref{fig:imf} shows the result. Our survey is essentially complete to M-star masses below $\sim0.3$ M$_\odot$, down to substellar objects with $\sim50$ M$_J$.  Extrapolating over the entire mass range using the \citet{2003PASP..115..763C} and \citet{2001MNRAS.322..231K}  mass functions, the total number of members in the cluster is $\approx$2\,600. This suggests that we are missing only $\approx$500 members, most of them massive and beyond the saturation limit of the current data. This value should be regarded as preliminary since a) the current data do not probe the entire area covered by the various groups of OriA-foreground; and b) the member selection suffers from some level of contamination, albeit estimated to be small ($\sim2$\%). The OriA-foreground population, or Blaauw's OB 1c subgroup, cannot be neglected in understanding the star formation history of the region, and perhaps even the ongoing star formation in Orion A, as we discuss in the next section.

\section{Discussion}

The present analysis confirms the results of \citet{2012A&A...547A..97A} and demonstrates that a large population  ($\ga$2600~sources)  of 5$\sim$10~Myr stars lies in front of the better known younger Orion A embedded population \citep[e.g.][]{2000AJ....120.3162L,2012AJ....144..192M}.
This foreground population is not uniformly distributed, but clusters loosely around
the NGC~1980 and NGC~1981 clusters, and possibly also around
the recently discovered L1641W \citep{2012A&A...547A..97A} and the newly found OriA-Fore~1 group.

The main result of this paper, the confirmation of a rich $\sim5$ Myr foreground population in the immediate vicinity of the Orion A cloud, fits the general scenario first proposed by \citet{1964ARA&A...2..213B} and refined by \citet{1977ApJ...214..725E} well, where the formation of stars in large complexes is not a  continuous process, but proceeds sequentially. As already pointed out in \citet{2012A&A...547A..97A}, this foreground population is not the optical counterpart of the embedded population in Orion A, but a distinct population perhaps responsible for the current star fomation in the integral-shaped filament in Orion A \citep{2008hsf1.book..459B}. 

As originally proposed by \citet{1964ARA&A...2..213B}, the Orion~OB1 association is
the product of several sequential star formation events, propagating
from the northwest, where the OB1a subgroup is located, towards the
belt region (OB1b), and then south towards OB1c (which includes NGC~1980 and NGC~1981), and more recently back towards Orion~A in the south,
and Orion~B east of the Belt stars, the OB1d subgroup \citep[see ][ for a review]{2008hsf1.book..459B}. The results presented
here suggest that while
the overall picture of sequential star formation is not invalidated by
our results, there is a higher level of complexity as more star
formation subunits present in the region are being discovered. The
most intriguing result seems to be that the age of NGC~1980 and NGC~1981,
separated by at least 14~pc, is not only indistinguishable within the
errors, but is also similar to that of the $\lambda$-Ori cluster,
located at least 100~pc away. At the same time, and also within the errors,
the distance to NGC~1980 and 1981 -- or the OB1c subgroup studied
here -- is similar to the distance of  25-Ori, $\sigma$-Ori,
$\lambda$-Ori, and Collinder~70 clusters, all located at about 380--385~pc (Table~\ref{tab:groups}) and spread across about 100~pc.  It is too early to speculate about
a 100~pc triggering or synchronization event in Orion, in particular
as 25-Ori and OB1a appear to be older than OB1b \citep[e.g ][]{2007ApJ...661.1119B}. But it is fair to say that it is not yet clear what the
casual relation is between all these star formation events with age
differences of only a few to $\sim$5~Myr but spread across about 100~pc of space. Better data are obviously needed to derive better ages and
parallaxes, proper motions, and radial velocities to attempt a more refined scenario for the star formation history of the Orion complex.

Regardless of the precise star formation history of Orion, and assuming a
normal IMF, a relatively large number of massive stars was formed in
the past 10~Myr. Several of these must have already exploded as
supernovae, which together with the feedback from the stellar winds
from existing stars are shaping the structure of the interstellar
medium in the region, including powering the large Barnard’s loop
\citep[][ see Fig.~\ref{fig:groups}]{2011ApJ...733....9O}. This feedback process probably
compressed the Orion~A molecular cloud (located 10--40 pc behind) and
triggered the ongoing star formation in the Orion Molecular Cloud,
including the ONC, L1641N, and OMC2/3, but also NGC~2024 (and possibly
L1641W and OriA-Fore 1).

\begin{figure*}
   \centering
   \includegraphics[width=0.95\textwidth]{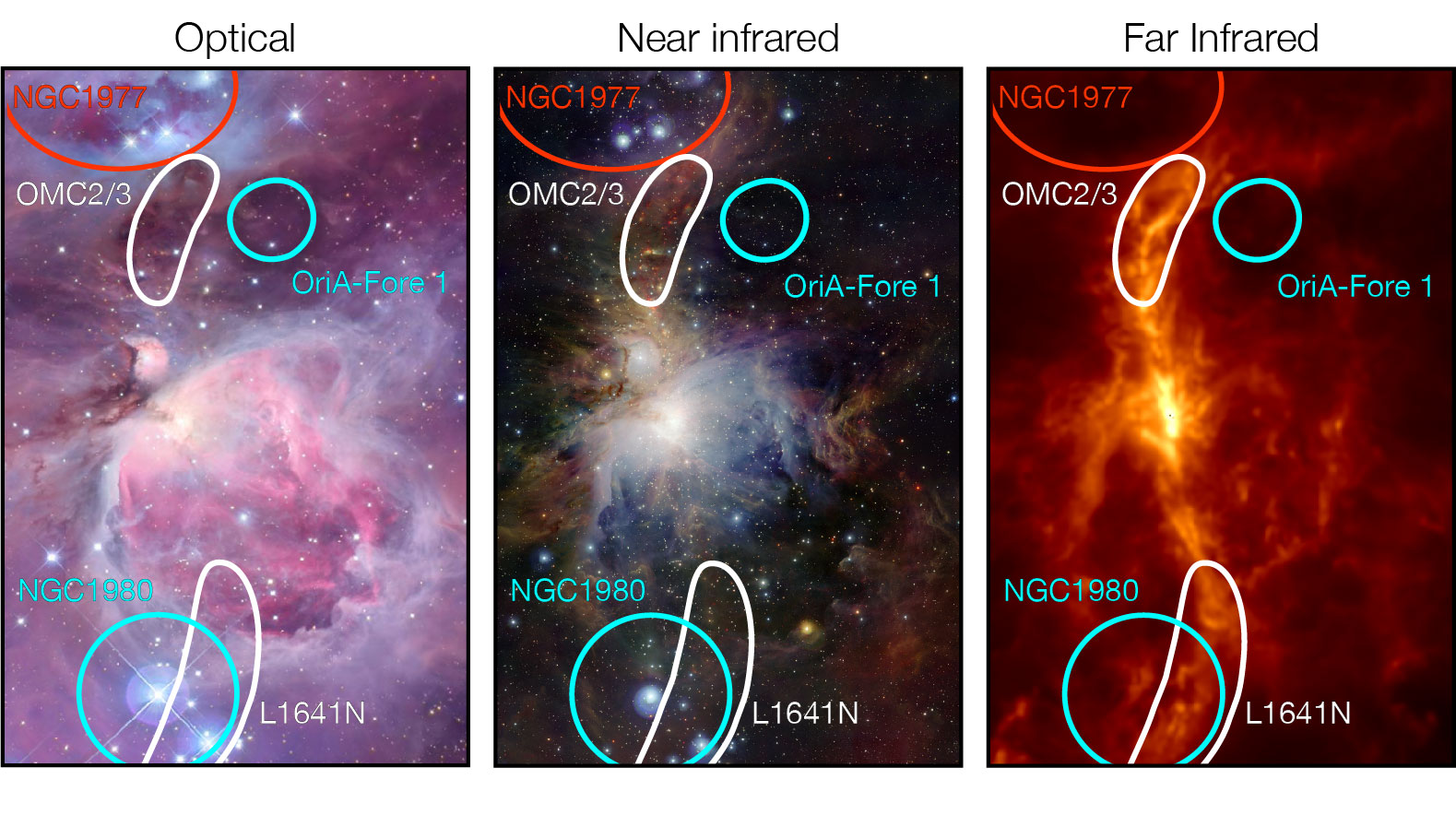}
      \caption{From left to right: optical, near-IR, and far-IR image of the Orion A cloud. Foreground groups are indicated in cyan. Groups associated to Orion~A are indicated in white. Group in the back of Orion~A are indicated in red. Credits:  Jon Christensen (optical); ESO/J. Emerson/VISTA/CASU (near-IR); H. Bouy (Far-IR Herschel SPIRE 500~$\mu$m).}
         \label{fig:integral}
   \end{figure*}

\begin{figure*}
   \centering
   \includegraphics[width=0.95\textwidth]{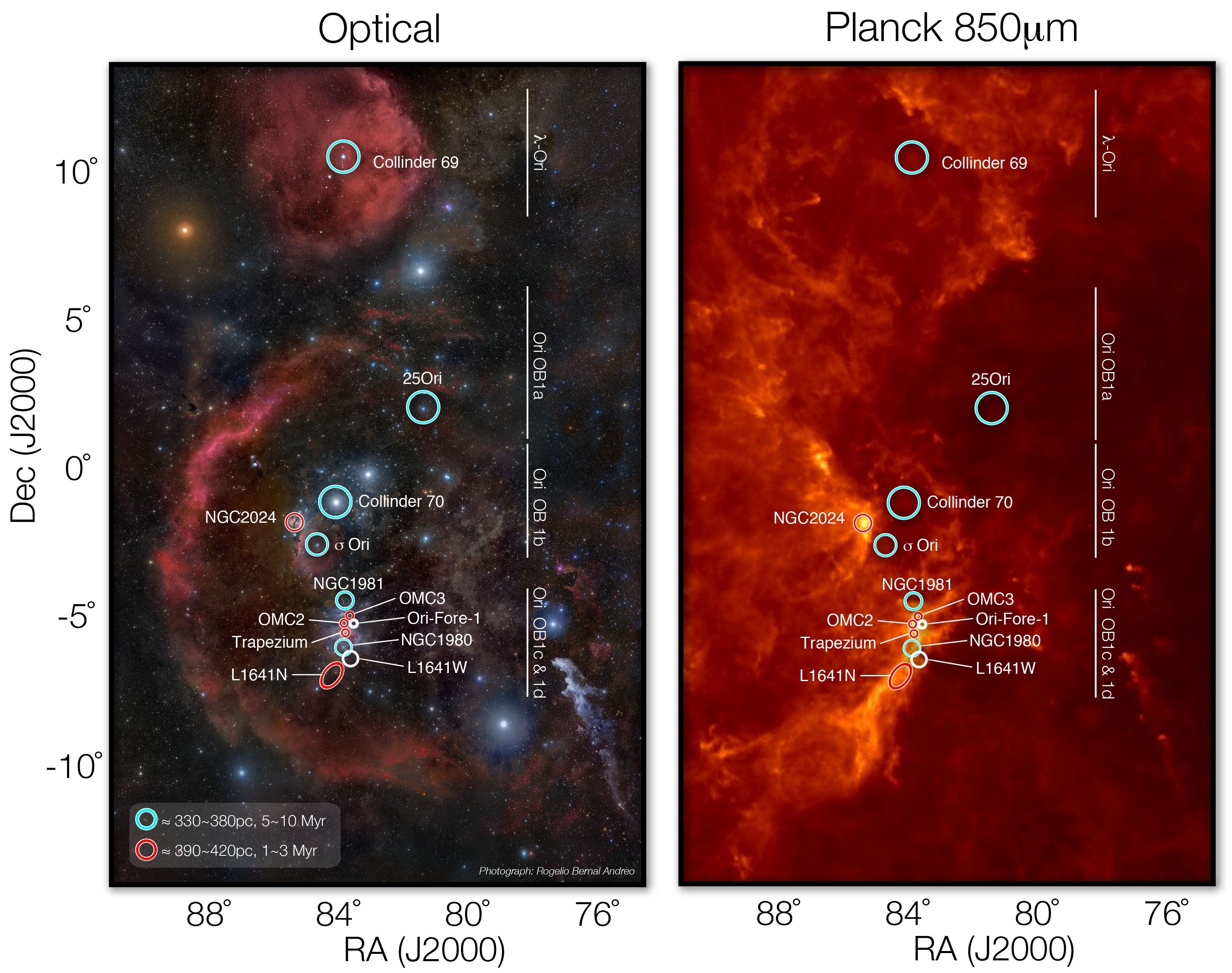}
      \caption{Left panel: Distribution of foreground (light blue) and background (red) groups overplotted on an optical photograph of the Orion constellation (courtesy of Rogelio Bernal Andreo - \emph{DeepSkyColors.com}). Right panel: Same as left panel, but overplotted on a far-infrared (850~$\mu$m) {\it Planck} map. L1641W and OriA-Fore~1 distances are uncertain and are represented as a white circles.}
         \label{fig:groups}
   \end{figure*}

Figure~\ref{fig:integral} shows the location of OriA-Fore~1 with respect to the nebulosities and to the cloud. Interestingly, the new group is located inside the northern elbow of the integral-shaped filament, in a region mostly free of dust as probed by {\it Herschel}, and free of HII nebulosities as probed by the optical. We speculate that the cloud might be squeezed between the massive cluster NGC~1977 in the background, and by the less massive OriA-Fore~1 in the foreground. The compression of the filament between these two groups results in the active star formation in OMC2 and OMC3. Similarly, the strong winds produced by the massive stars of NGC~1980 might be giving its shape to the southern elbow of the integral-shaped filament, and be triggering the star formation in L1641N. 

Figure~\ref{fig:groups} (right panel) shows a schematic representation of the spatial distribution of the various groups on top of a far-infrared map probing the molecular clouds.  The most distant and youngest groups all closely follow the cloud. On the other hand, all the closest and older groups but NGC~1980 are located systematically away from the cloud, providing additional evidence that they are not the optical counterpart of the embedded population, but a distinct population, corresponding to a previous star formation episode. The fact that the foreground population to Orion A characterized in this paper generally aligns with the orientation of the Orion A cloud is puzzling and may provide information about the original distribution of molecular gas, but can also be interpreted as a coincidence -  surely the reason why so little attention has been dedicated in the literature to this important population.

This scenario provides a simple explanation for the age spread observed in the region \citep[see e.g ][and references therein]{2006ApJ...644..355H}. A fraction of the spread might be purely observational and caused
by the contamination from the older coincident foreground population. But most of the age spread is probably real and the result of  successively triggered star formation events and consequent episodic bursts of the star formation rate.

\section{Conclusions}
We combined new wide-field optical observations of the Orion~A group of associations with archival data and catalogs. A dense sequence appears clearly in all color-magnitude diagrams. The observed match with empirical sequences suggests an age of 5$\sim$10~Myr and a distance of $\approx$380~pc, placing the corresponding sources in the immediate front of the Orion A cloud and the ONC. Members of the sequence were selected using a novel multi-dimensional probabilistic analysis using carefully selected color-magnitude diagrams and luminosities. The sources cluster into several groups, which can be spatially associated to NGC~1980, L1641N, L1641W, and NGC~1981, and a new group OriA-Fore~1 located southwest of OMC3. The current work confirms the results reported in \citet{2012A&A...547A..97A} using X-ray, optical, and infrared data.

We propose the following scenario to explain the observations in the region: a first generation of clusters and associations located at 380$\sim$400~pc formed between 5$\sim$10~Myr ago. It included NGC~1980, NGC~1981, 25-Ori, $\sigma-$Ori, $\lambda-$Ori, and possibly L1641W and OriA-Fore~1.  These groups formed several thousands of stars, including a few massive stars. After a few million years, the most massive stars must have exploded as supernovae, which together with stellar winds very likely triggered episodic star formation events in the ONC, resulting in the Trapezium, L1641N, OMC2/3, and maybe even NGC~2024 in Orion B. 

\acknowledgements We are grateful to Rogelio Bernal Andreo and Jon Christensen for granting us permission to use their  photographs of Orion. H. Bouy is funded by the Spanish Ram\'on y Cajal fellowship program number RYC-2009-04497. We acknowledge support from the Faculty of the European Space Astronomy Centre (ESAC). This research has been funded by Spanish grants AYA2010-21161-C02-02, AYA2012-38897-C02-01. This publication is supported by the Austrian Science Fund (FWF). We are grateful to Amelia Bayo for providing us with an electronic version of the Collinder~69 catalog. This research made use of Astropy, a community-developed core Python package for Astronomy \citep{2013A&A...558A..33A}.
This work has made an extensive use of Topcat \citep[\url{http://www.star.bristol.ac.uk/~mbt/topcat/},][]{2005ASPC..347...29T}. This research has made use of the VizieR and Aladin images and catalogue access tools and of the SIMBAD database, operated at CDS, Strasbourg, France. This research has made use of VOSA and the Spanish Virtual Observatory.
Based on observations obtained with MegaPrime/MegaCam, a joint project of CFHT and CEA/DAPNIA, at the Canada-France-Hawaii Telescope (CFHT) which is operated by the National Research Council (NRC) of Canada, the Institute National des Sciences de l'Univers of the Centre National de la Recherche Scientifique of France, and the University of Hawaii. 
This research has made use of the APASS database, located at the AAVSO web site. Funding for APASS has been provided by the Robert Martin Ayers Sciences Fund. We are grateful to the AAVSO team for giving us access to the APASS DR1 catalogue. 
This research is based on data obtained at the Cerro Tololo Inter-American Observatory, National Optical Astronomy Observatory, which are operated by the Association of Universities for Research in Astronomy, under contract with the National Science Foundation.
This work is based in part on data obtained as part of the UKIRT Infrared Deep Sky Survey.
This research made use of the SDSS-III catalogue. Funding for SDSS-III has been provided by the Alfred P. Sloan Foundation, the Participating Institutions, the National Science Foundation, and the U.S. Department of Energy Office of Science. The SDSS-III web site is http://www.sdss3.org/. SDSS-III is managed by the Astrophysical Research Consortium for the Participating Institutions of the SDSS-III Collaboration including the University of Arizona, the Brazilian Participation Group, Brookhaven National Laboratory, University of Cambridge, University of Florida, the French Participation Group, the German Participation Group, the Instituto de Astrofisica de Canarias, the Michigan State/Notre Dame/JINA Participation Group, Johns Hopkins University, Lawrence Berkeley National Laboratory, Max Planck Institute for Astrophysics, New Mexico State University, New York University, Ohio State University, Pennsylvania State University, University of Portsmouth, Princeton University, the Spanish Participation Group, University of Tokyo, University of Utah, Vanderbilt University, University of Virginia, University of Washington, and Yale University.

This publication makes use of data products from the Wide-field Infrared Survey Explorer, which is a joint project of the University of California, Los Angeles, and the Jet Propulsion Laboratory/California Institute of Technology, funded by the National Aeronautics and Space Administration.
This research used the facilities of the Canadian Astronomy Data Centre operated by the National Research Council of Canada with the support of the Canadian Space Agency.   
This publication makes use of data products from the Two Micron All Sky Survey, which is a joint project of the University of Massachusetts and the Infrared Processing and Analysis Center/California Institute of Technology, funded by the National Aeronautics and Space Administration and the National Science Foundation.
This work is based in part on observations made with the Spitzer Space Telescope, which is operated by the Jet Propulsion Laboratory, California Institute of Technology under a contract with NASA.

\bibliographystyle{aa}

\Online

\begin{appendix} 

\section{Optical and near-infrared photometry}

\begin{deluxetable}{lccccccccccc}
\rotate
\tabletypesize{\scriptsize}
\tablecaption{Astrometry, photometry and membership probability to NGC~1980 \label{tab:catalogue}}
\tablewidth{0pt}
\tablehead{
\colhead{ID} & \colhead{R.A} & \colhead{Dec} & \colhead{$g$} & \colhead{$r$} & \colhead{$i$} & \colhead{$z$} & \colhead{$Y$} & \colhead{$J$} & \colhead{$H$} & \colhead{$K$} & \colhead{Proba} \\
\colhead{} & \colhead{(J2000)} & \colhead{(J2000)} &  \colhead{(mag)} &   \colhead{(mag)} &   \colhead{(mag)} &   \colhead{(mag)} &   \colhead{(mag)} &   \colhead{(mag)} &   \colhead{(mag)} &   \colhead{(mag)} &   \colhead{(\%)} 
}
\startdata
1 & 82.215192 & -5.46041214899587 & 17.85$\pm$0.06 &     &     &     &     &     &     &     &  \\
2 & 82.875308 & -5.53696629695448 & 21.43$\pm$0.19 &     &     &     &     &     &     &     &  \\
3 & 82.177325 & -5.73258047115825 & 21.54$\pm$0.21 &     &     &     &     &     &     &     &  \\
4 & 82.884495 & -5.69717042076206 & 16.46$\pm$0.06 &     &     &     &     &     &     &     &  \\
5 & 82.871700 & -5.31117868183994 & 21.59$\pm$0.19 &     &     &     &     &     &     &     &  \\
\nodata & \nodata & \nodata & \nodata & \nodata & \nodata & \nodata & \nodata & \nodata & \nodata & \nodata & \nodata \\
318195 & 82.8890372 & -6.1383945 &     &     & 22.3$\pm$0.08 & 21.78$\pm$0.09 &     &     &     &     & 0.0 \\
318196 & 82.8530871 & -6.1373586 &     &     & 22.44$\pm$0.06 & 21.89$\pm$0.08 &     &     &     &     & 0.0 \\
318197 & 82.7754450 & -6.1370855 &     &     &     & 19.75$\pm$0.03 &     &     &     &     & 0.0 \\
318198 & 82.7844997 & -6.1369484 & 23.18$\pm$0.13 &     & 21.62$\pm$0.04 & 21.36$\pm$0.07 &     &     &     &     & 0.0 \\
318199 & 82.6534512 & -6.1362583 &     &     & 21.31$\pm$0.03 & 20.68$\pm$0.03 & 19.82$\pm$0.16 & 19.35$\pm$0.19 &     &     & 0.0 \\
318200 & 82.8885940 & -6.1364786 &     &     & 21.79$\pm$0.04 & 21.27$\pm$0.05 &     &     &     &     & 0.0 \\
318201 & 82.8074845 & -6.1360497 &     &     & 22.73$\pm$0.09 & 22.04$\pm$0.11 &     &     &     &     & 0.0 \\
318202 & 82.8105568 & -6.1356670 &     &     & 22.1$\pm$0.05 & 21.24$\pm$0.06 &     &     &     &     & 0.0 \\
318203 & 82.7055493 & -6.1350536 &     &     & 22.34$\pm$0.05 & 21.6$\pm$0.07 &     &     &     &     & 0.0 \\
318204 & 82.8436510 & -6.1352804 &     &     & 21.36$\pm$0.03 & 20.66$\pm$0.03 &     & 19.12$\pm$0.15 & 18.24$\pm$0.18 &     & 0.0 \\
318205 & 82.6438898 & -6.1347576 & 23.77$\pm$0.18 &     & 20.73$\pm$0.03 & 20.8$\pm$0.04 & 20.23$\pm$0.23 &     &     & 17.95$\pm$0.17 & 0.0 \\
\nodata & \nodata & \nodata & \nodata & \nodata & \nodata & \nodata & \nodata & \nodata & \nodata & \nodata & \nodata \\
132153 & 84.2702543 & -4.2309804 & 21.39$\pm$0.05 & 19.81$\pm$0.02 & 17.69$\pm$0.01 & 16.49$\pm$0.01 & 15.28$\pm$0.0 & 14.62$\pm$0.0 & 14.12$\pm$0.0 &     & 100.0 \\
132154 & 84.2739473 & -4.7892626 & 18.15$\pm$0.01 & 16.73$\pm$0.01 & 15.38$\pm$0.01 & 14.59$\pm$0.0 &     &     &     &     & 100.0 \\
132155 & 84.2528004 & -5.4169561  & 17.38$\pm$0.01 & 15.99$\pm$0.01 & 14.54$\pm$0.01 & 13.75$\pm$0.0 &     &     &     &     & 100.0 \\
132156 & 83.6885989 & -5.2738119 &     & 19.63$\pm$0.06 & 17.24$\pm$0.02 &     & 14.77$\pm$0.0 & 14.16$\pm$0.0 & 13.53$\pm$0.04 & 13.23$\pm$0.0 & 99.9 \\
132157 & 84.2981832 & -6.1125390 &     & 18.37$\pm$0.05 & 16.49$\pm$0.02 &     &     & 14.04$\pm$0.03 & 13.48$\pm$0.0 & 13.17$\pm$0.04 & 99.9 \\
132158 & 84.1374590 & -6.0416210 & 18.3$\pm$0.07 & 16.93$\pm$0.05 & 15.49$\pm$0.02 &     &     & 13.42$\pm$0.03 & 12.73$\pm$0.0 & 12.48$\pm$0.0 & 99.9 \\
132159 & 83.7844179 & -5.4866692 & 19.69$\pm$0.1 & 18.6$\pm$0.05 & 16.57$\pm$0.02 &     & 14.5$\pm$0.02 & 13.82$\pm$0.04 & 13.06$\pm$0.03 & 12.67$\pm$0.02 & 99.9 \\
\enddata
\end{deluxetable}

\end{appendix}

\end{document}